\documentclass{ws-p10x7}
\newcommand{\DO}{$ {\rm D}_2 {\rm O}$}
\newcommand{\HO}{$ {\rm H}_2 {\rm O}$}
\newcommand{\NS}{$^{16}{\rm N}$}
\newcommand{\Li}{$^{8}{\rm Li}$}

\begin{document}

\title{Solar Neutrino Results from the Sudbury Neutrino Observatory}

\author{Joshua R. Klein, for the SNO Collaboration}

\address{
Department of Physics and Astronomy, University of Pennsylvania,
Philadelphia, PA 19104-6396 \\
}

\twocolumn[\maketitle\abstract{We describe here the measurement of the
flux of neutrinos created by the decay of solar $^8$B by the
Sudbury Neutrino Observatory (SNO).  The neutrinos were detected
via the charged current (CC) reaction on deuterium and by the elastic 
scattering (ES) of electrons.  The CC reaction is sensitive
 exclusively to $\nu_e$'s, while the ES reaction also has
 a small sensitivity to $\nu_{\mu}$'s and $\nu_{\tau}$'s.
 The flux of $\nu_e$'s from $^8$B decay measured by
 the CC reaction rate is  $\phi^{\rm CC}(\nu_e) = 1.75 \pm 0.07~({\rm stat.})^{+0.1
2}_{-0.11}~({\rm sys.}) \pm 0.05~({\rm theor.}) \times 10^6~{\rm cm}^{-2} {\rm s}^{-1}$. Assuming no flavor transformation,  the flux inferred from the ES
 reaction rate is  $\phi^{\rm ES}(\nu_x)=2.39\pm 0.34~({\rm stat.}) ^{+0.16}_{-0.14}~({\rm sys.}) \times 10^6~{\rm cm}^{-2} {\rm s}^{-1}$.
 Comparison of $\phi^{\rm CC}(\nu_e)$ to the Super-Kamiokande Collaboration's
 precision  value of $\phi^{\rm ES}(\nu_x)$  yields a $3.3\sigma$ difference,
assuming the systematic uncertainties are normally distributed,
 providing evidence that there is a non-electron flavor active
 neutrino component in the solar flux.  The total flux of active  $^8$B
neutrinos is thus determined to be
 $5.44\pm 0.99 \times 10^6~{\rm cm}^{-2} {\rm s}^{-1}$,
in close agreement with the predictions of solar models. 
}]

\section{Introduction}

	Over thirty years of solar neutrino 
experiments~\cite{cl,kam,sage,gallex,SK,gno} have demonstrated that the flux
of neutrinos from all sources within the Sun is significantly smaller 
than predicted by models of the Sun's energy generating 
mechanisms~\cite{BPB,TC}.  The
deficit is not only universally observed but has an energy dependence which
makes it hard to attribute to astrophysical sources: the data are consistent
with a negligible flux of neutrinos from solar $^7$Be~\cite{hata,hamish}, 
though neutrinos
from $^8$B (a product of solar $^7$Be reactions) are observed.  A
natural explanation for the observations is that neutrinos born as
$\nu_e$'s change flavor on their way to the Earth, thus producing an 
apparent deficit in experiments detecting primarily $\nu_e$'s.
Neutrino oscillations---either in vacuum or matter---provide a mechanism
both for the flavor change and the observed energy variations.

	While these deficit measurements argue strongly for  neutrino flavor 
change through oscillation, a far more compelling demonstration would not
resort to model predictions at all but look for non-$\nu_e$ flavors coming 
from the Sun.  The Sudbury Neutrino Observatory (SNO) was designed to do just 
that: provide direct evidence of solar neutrino flavor change through 
the inclusive appearance of non-electron neutrino flavors from the Sun.  
We present here the first solar neutrino results from SNO, which have
also been described in an earlier publication~\cite{snoprl}.

\section{SNO Detector}

\begin{figure}
\epsfxsize130pt
\figurebox{130pt}{180pt}{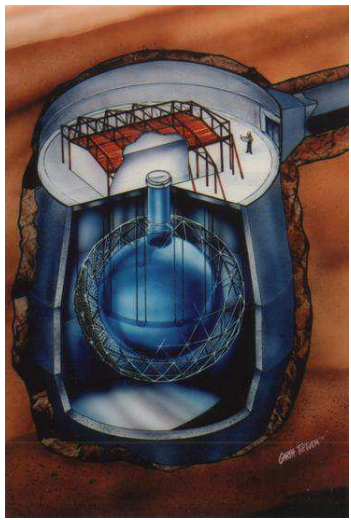}
\caption{Diagram of SNO Detector.\label{fig:detector}}
\end{figure}

	SNO is an imaging water Cerenkov detector, which uses heavy
water (D$_2$O) as both the interaction and detection medium~\cite{NIM}.  
Figure~\ref{fig:detector} shows a diagram of the detector.  SNO
is located $\sim$ 2~km (6020 k.w.e.) underground in INCO Ltd.'s 
Creighton Mine,  deep enough that the rate of cosmic ray muons passing
through the entire active volume is just 3/hour. 

The 1000 tons of heavy water
is contained in a 12~m diameter transparent acrylic vessel, and is 
surrounded by 2 ktons of light water shielding.  The Cerenkov light produced 
by neutrinos and radioactive backgrounds is detected by an array of 9500
8 inch photomultiplier tubes (PMTs), supported by a stainless steel geodesic
sphere.  Each PMT is surrounded by a light concentrator, which increases
the photocathode coverage to nearly $\sim 55$\%.  The  front-end discriminator 
thresholds are set to fire on 1/4 of a photoelectron of charge.  Outside the 
PMT support sphere is another 7 ktons of light water shielding.

	The detector is also equipped with a flexible calibration system,
capable of placing sources almost everywhere in either the $x-z$ or $y-z$
plane.  The sources that can be deployed include a diffuse multi-wavelength
laser for measurements of optical parameters and PMT timing, a
$^{16}$N source which provides a triggered sample of 6.13~MeV $\gamma$'s,
and a $^8$Li source delivering $\beta$'s with an endpoint near 14~MeV.
In addition, high energy energy (19.8~MeV) $\gamma$'s are provided
by a $^3{\rm H}(p,\gamma)^4{\rm He}$ (`pT') source~\cite{poon} and neutrons
by a Cf source.  Some of the sources can also be deployed on vertical axes 
within the light water volume between the acrylic vessel and PMT support sphere.

\section{SNO Reactions}

	SNO can provide direct evidence of solar neutrino flavor
change through comparisons of the interaction rates of three different
processes:

\begin{center}
  \begin{tabular}{ll}
     $ \nu_x + e^- \rightarrow \nu_x + e^-$  & (ES)\\
     $\nu_e + d \rightarrow p + p + e^-$\hspace{0.5in} & (CC)\\
     $ \nu_x + d \rightarrow p + n + \nu_x$ & (NC)\\  \\
  \end{tabular}
 \end{center}

	The first reaction, the elastic scattering (ES) of electrons,
has been used to detect solar neutrinos in other water Cerenkov experiments.
It has the great advantage that the recoil electron direction is strongly
correlated with the direction of the incident neutrino (and hence the
direction of the Sun).  In addition, this reaction has sensitivity 
to all neutrino flavors.  For $\nu_e$'s, the elastic scattering reaction has 
both charged and neutral current components,  making the cross section for 
$\nu_e$'s 6.5 times larger than that for $\nu_{\mu}$'s  or $\nu_{\tau}$'s.

	The deuterium in the heavy water makes the second process possible:  
an exclusively charged current (CC) reaction which (at solar energies) occurs 
only for $\nu_e$'s.  In addition to providing exclusive sensitivity to 
$\nu_e$'s, this reaction has the advantage that the recoil electron
energy is strongly correlated with the incident neutrino energy, and thus
can provide a good measurement of the $^8$B energy spectrum. The CC reaction
also has an angular correlation with the Sun which falls as
$(1-0.340\cos(\theta_{\odot}))$~\cite{vogel}, and has a much larger
cross section ($\sim$ 10 times larger) than the ES reaction.

	The third reaction---also unique to heavy water---is a purely
neutral current process.  This has the obvious advantage that it is equally
sensitive to all neutrino flavors, and thus provides a direct 
model-independent measurement of the total flux of neutrinos from the Sun.

	For both the ES and CC reactions, the recoil electrons are 
directly detected through their production of Cerenkov light.  For the NC
reaction, the neutrons are not seen directly, but are detected in a 
multi-step process.  When a neutrino liberates a neutron from a deuteron,
the neutron wanders within the D$_2$O and is eventually captured by another
deuteron, releasing a 6.25~MeV $\gamma$ ray.  The $\gamma$ Compton scatters
an electron and it is this secondary particle which is detected.  

Although the data we present here were acquired with the acrylic vessel filled 
with pure D$_2$O, the detector is now running with NaCl added to the heavy 
water.  The addition of the salt provides chlorine which has a larger capture
cross section (and hence a higher detection efficiency) for the neutrons.
The capture on chlorine also yields multiple $\gamma$'s instead of the single
$\gamma$ from the pure D$_2$O phase, which aids in the identification of
neutron events.  Eventually, discrete He$^3$ proportional
counters will be added which will count neutrons exclusively.

	To determine whether neutrinos which start out as $\nu_e$'s in the
solar core convert to another flavor before detection on Earth, we have
two choices: comparison of the CC reaction rate to the NC reaction rate,
or comparison of the CC rate to the ES rate.  The former has the advantage
of high sensitivity---we compare the total flux to the $\nu_e$ flux and
therefore expect to see a large difference if the true neutrino flux
agrees with standard solar models (which predict a total flux two to three
times larger than previous measurements).  In addition, uncertainties in the
cross sections for the two processes will largely cancel.  

	The second comparison has the advantage that both the CC and ES recoil 
electrons provide neutrino spectral information.  The spectral information can 
ultimately be used to show that any excess in the ES reaction over the CC 
reaction is not caused by a difference in the energy thresholds used to analyze
the two reactions.  The CC-ES comparison  also has the advantage that the
strong angular correlation with the Sun demonstrates that any excess seen is
not due to some unexpected non-solar background.  Lastly, the CC-ES 
comparison can be made with fairly high precision despite the small ES
reaction cross section, because the Super-Kamiokande collaboration has
already made a precision, high statistics measurement of the ES rate~\cite{SK}.
For the results presented here, only the CC-ES comparison will be described.

\section{Data Analysis}

	The goal of the data analysis is the determination of the 
relative sizes of the three signals (CC, ES, and neutrons) and ultimately
the comparison of the rates.  In the pure D$_2$O detector 
configuration---the configuration with which these data were taken---we 
cannot separate the signals on an event-by-event basis.  Instead, we
`extract' the signals statistically by using the fact that they are
distributed distinctly in the following three derived quantities: the
kinetic energy of the recoil electron or capture $\gamma$ ray (T), the 
reconstructed radial position of the interaction ($R^3$), and the reconstructed 
direction of the event relative to the Sun ($\cos \theta_{\odot}$). 

\begin{figure}
\epsfxsize15pc
\figurebox{16pc}{32pc}{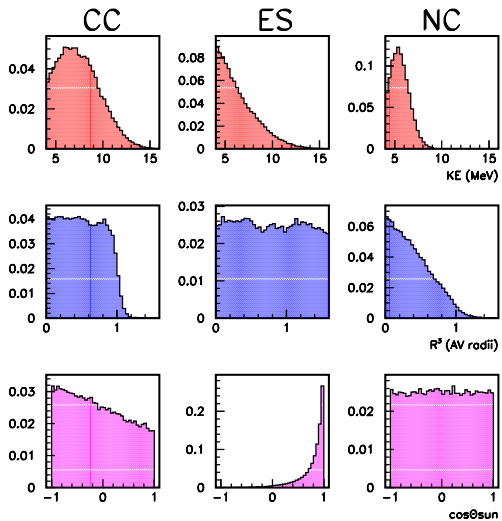}
\caption{The energy (top), radial (middle), and directional (bottom) 
distributions used to build pdfs to fit the SNO signal data. \label{fig:pdfs}}
\end{figure}     
	Figure~\ref{fig:pdfs} shows these distributions for each of the
signals.   The top row of Figure~\ref{fig:pdfs} plots the different
energy distributions for the three signals.  We see in the figure that
the strong correlation between the electron energy and the incident neutrino
energy for the CC interaction produces a spectrum which resembles the initial 
$^8$B neutrino spectrum, while the recoil spectrum for the ES reaction is
much softer.  The NC reaction is---within the resolution smearing
of the detector---essentially a $\delta$-function, because the $\gamma$
produced by the neutron capture on deuterium always has the same 6.25~MeV.

	The bottom row of Figure~\ref{fig:pdfs} shows the reconstructed
direction distribution of the events.  In the middle of that row we see
the familiar peaking for the ES reaction, pointing toward the Sun.  The
$\sim$ $1-1/3\cos \theta_{\odot}$ distribution of the CC reaction is also 
clear in the left hand side of the bottom row.  Not surprisingly, the NC 
reaction shows no correlation with the solar direction---the $\gamma$ ray
from the captured neutron knows nothing about the incident neutrino.

	The distributions of reconstructed event positions is shown in
the middle row of Figure~\ref{fig:pdfs}. These distributions are 
plotted as a function of $R^3$, with $R^3=1$ occurring at the radius of the
acrylic vessel (the edge of heavy water volume).  We see here that the
CC reactions---which occur only on deuterons---produce events distributed 
uniformly within the heavy water, while the ES reaction (which occurs on
any electron) produces events distributed uniformly well beyond the
heavy water volume.  The small leakage of events just outside the heavy
water volume for the CC reaction is due to the resolution tail of the 
reconstruction algorithm.  

The NC reaction, however, does not have a uniform 
distribution inside the heavy water like the CC reaction, but instead 
monotonically decreases from the central region to the edge of the acrylic 
vessel.  The reason for this is that the capture cross section for neutrons on
deuterium is very small---the neutron wanders around long enough inside
the D$_2$O that it may leak outside and be captured by hydrogen in either the 
acrylic vessel or the light water.  Such hydrogen captures produce
a much lower energy $\gamma$ ray ($\sim$ 2.2~MeV), below the analysis threshold.
Therefore the acceptance for events which are produced near the edge of the 
volume is reduced, because the probability of leakage there is
correspondingly higher than for events produced near the center.

	One last point needs to be made regarding the distributions
labelled `NC' in Figure~\ref{fig:pdfs}: they represent equally well
the detector response to all neutrons, not just those produced by 
neutral current interactions, as long as the neutrons are produced 
uniformly in the detector.  For example, neutrons produced through  
photodisintegration by $\gamma$ rays emitted by U or Th chain daughters
inside the D$_2$O will have the same distributions of energy, radial
position, and solar direction as those produced by solar neutrinos.
In the analysis described here, no separation is done between these neutrons
and those from the NC reaction.

	To determine the size of the three signals, then, we use
these nine distributions to create probability density functions (pdfs)
and perform a generalized maximum likelihood fit to the same distributions
in the data.  There are, however, two principal prerequisites that must be
satisfied before we can even begin this `signal extraction' process.  First,
we need to process the data so that it is in a form we can use to do the
fits.  For example, we need to reconstruct the events to give us positions and 
directions that can be used to produce distributions, and we need to
calibrate the energy of each event.  Even more importantly, we must 
be sure that the {\it only} signals present in the data are the
three for which we are doing the fits---we have implicitly assumed that the
backgrounds are negligible.  To accomplish this, we 
apply cuts to the data to eliminate backgrounds and we must ultimately 
demonstrate that any residual backgrounds are negligible.

	The second signal extraction prerequisite is that the distributions 
used in the fitting process must be good representations
of the detector's true response.  In other words, we must build a model
of the detector's response to the three signals which can be used to
generate the pdfs used in the fit.  The model needs to reproduce the
response at all places in the detector, for all neutrino directions, for
all neutrino energies, and for all times.  The last requirement is necessary
because the detector's response changes over time due to things like
failed PMT's or electronics channels.

	The analysis we describe here therefore has three major components
before the final fitting stage: the processing of data to remove backgrounds, 
the building of a model to fit the data, and the demonstration that the 
residual backgrounds are small enough to use the signal-only model in the fits.

\subsection{Data Processing}
\label{sec:dproc}

	We recorded the data set used in this analysis between 
November 2, 1999, and January 15, 2000.  Roughly 40\% of the time during
this period was taken up either by calibration source runs or downtime caused
by mine power outages.  Of the remaining good data, we selected runs to analyze
based on criteria which were `blind' to the data itself---whether enough
channels were live, whether calibration sources were present, whether water
assays were being run, etc.  After passing this run selection stage, no
further run removal was allowed from the data set, and the final total livetime
amounted to 241 days.  Approximately 30\% of the data was put aside
to serve as a blind test of statistical bias.  As no significant differences
were found between this sample and the other 70\%, all subsequent discussion
here refers to the full data set.

	During this time, the primary trigger threshold was set 
to fire on a $\sim$ 100~ns coincidence of 18 PMT's each exceeding a channel
threshold of $\sim$ 1/4 photoelectron. This trigger threshold corresponds
to an energy of roughly 2~MeV.  The trigger reaches 100\% efficiency at 
23  hit PMTs.  In addition to the 100~ns coincidence, we ran simultaneously 
with other triggers, such as a pre-scaled (1:1000) lower threshold (11 PMTs) 
trigger, a trigger on PMT pulse height sums, and a pulsed (random) trigger.

\begin{figure}
\epsfxsize130pt
\figurebox{120pt}{160pt}{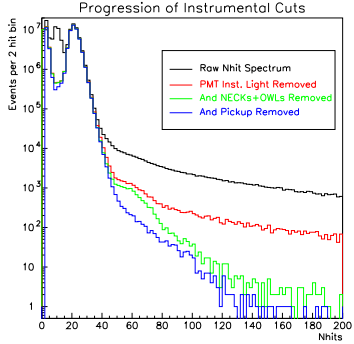}
\caption{Effects of progressive application of instrumental background
cuts.\label{fig:dcprog}}
\end{figure}

	The raw data set is far from the clean distributions shown
in Figure~\ref{fig:pdfs}.  In particular, the data is contaminated by 
instrumental backgrounds arising primarily from PMT light emission 
(`flasher PMTs'), static discharges in the neck of the acrylic vessel, 
or electronic pickup.  Although these instrumental backgrounds are very
distinct from the neutrino signal, they occur at far higher rates: flasher
events, for example, occur roughly once each minute compared to the
five to ten neutrino events we expect each day.  We therefore developed
a suite of low level cuts designed to remove instrumental backgrounds while 
losing a minimum of neutrino events.  These cuts were applied before any
reconstruction of the data was done, and used only primitive information 
such as the PMT charge distributions, the raw and calibrated time 
distributions, hits in veto tubes, and event-to-event time correlations.
Figure~\ref{fig:dcprog} shows the effects of the progressive application
of these instrumental background cuts to the raw data set, illustrating the
multiple orders of magnitude reduction in the overall number of events.

\begin{figure}
\epsfxsize130pt
\figurebox{120pt}{160pt}{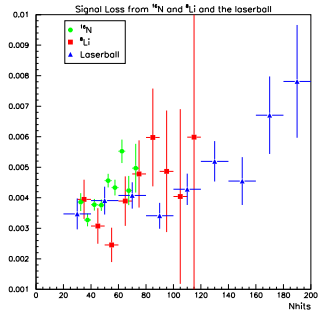}
\caption{Acceptance loss from low level instrumental background cuts, measured
with calibration sources.\label{fig:dcsac}}
\end{figure}

	In any case in which such a large reduction is obtained, the obvious
question is what is the consequent reduction in good events---how
much acceptance loss have we incurred by applying cuts which remove more than 
three orders of magnitude of the instrumental backgrounds?  To measure
this loss, we used triggered calibration sources which provided both samples of
Cerenkov events and isotropic light events, and applied the same cuts to
the source data as to the neutrino data.  Figure~\ref{fig:dcsac} shows
the acceptance loss as a function of the number of hit PMTs, for 
\NS, \Li, and laser data.  Although there is evidence of a bias at
high energies (high number of hit PMTs), the overall scale of the loss is 
very small, $\sim$ 0.5\%.

\begin{figure}
\epsfxsize130pt
\figurebox{120pt}{160pt}{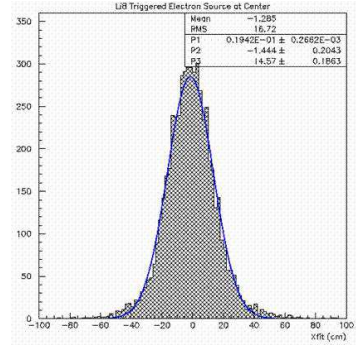}
\caption{Resolution in $x$ for \Li~source data. \label{fig:li8res}}
\end{figure}

	For events passing this first stage, we reconstructed the vertex
position and direction of the particle using the calibrated times and
positions of the hit PMTs.  The reconstruction algorithm begins
with maximum likelihood fits using only PMT times, seeded by positions
fixed to a grid throughout the detector volume.  The best fit vertex from
this grid-seeded procedure is then used as a  seed for a second level of 
fitting which uses both the PMT times and their angular distribution to 
simultaneously fit both the position and the direction of the event.  The 
fitting process includes cuts on angular figures-of-merit which test both the 
quality of the fit and the hypothesis that the event is a single Cerenkov 
electron.   Figure~\ref{fig:li8res} shows the vertex resolution for electrons 
produced by the \Li~source, which provides a localized ($\sim 5$~cm) set of 
electrons with a broad spectrum of energies.  At \NS~energies, 
the vertex resolution is 16~cm and the angular resolution is $26.7^{\circ}$.

	For each event surviving the reconstruction stage, we assigned
an energy based on the hypothesis that the event was a single Cerenkov electron.
While the number of hit PMTs by itself is directly related to the event 
energy, it must be corrected for the number of live channels online when the 
event was recorded, any change in the overall detector gain with time, and the 
optical effects of the intervening media between the Cerenkov production point 
(the event position) and the photon detection points (the hit PMTs).  The 
optical corrections were calculated using in-situ measurements of the 
detector's optical properties (attenuation lengths, PMT angular responses, 
etc.) and account for both the vertex position and the event direction.  To 
minimize uncertainties associated with late hits (reflections, scattering, 
noise), the optical corrections use only prompt (in-time) photons by requiring 
the fitted time residuals of the PMT hits to be within a narrow window around 
$\Delta t=0$~ns.

\begin{figure*}
\epsfxsize25pc
\figurebox{16pc}{32pc}{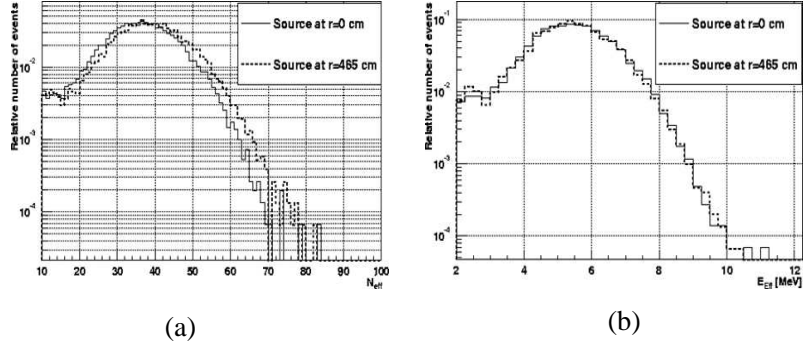}
\caption{Comparison of \NS~energy distribution before (a) and after (b)
optical corrections.\label{fig:ecal}}
\end{figure*}

Figure~\ref{fig:ecal} shows the calibrated response of the 
detector to \NS~data.  In Figure~\ref{fig:ecal}a we see the energy
distributions for data taken with the source at the center and at
$R=465$~cm. The only corrections made here are for the number of live channels
online, and therefore the shift in the mean of the two distributions is due to 
the different optical response at the two positions. Figure~\ref{fig:ecal}b
demonstrates how the two distributions coincide once the optical
corrections are applied.

\begin{figure}
\epsfxsize130pt
\figurebox{120pt}{160pt}{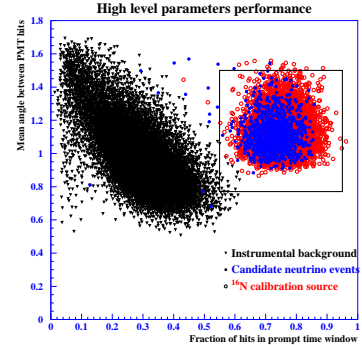}
\caption{Cerenkov box cuts defined by average angle between hit PMTs
and ratio of prompt to total light.\label{fig:cerbox}} 
\end{figure}

	With the event positions and directions fit and the energy
calibrated, we passed the data through a final stage of cuts aimed at 
ensuring that the remaining events were consistent with Cerenkov light.  
We defined Cerenkov light with two orthogonal cuts: one which tests the 
narrowness of the timing distribution, and one which tests the angular 
distribution of PMT hits.  The former is done by cutting on the ratio of 
prompt (in-time) hits to the total number of hits in the event, and the latter 
by using the average angular distance between hit PMTs in the event.
Figure~\ref{fig:cerbox} shows three data sets
distributed in these two variables: data tagged by the low level
instrumental background cuts (triangles), Cerenkov data from the
\NS~source (open circles), and neutrino data (closed circles).  The
box used to define Cerenkov light is also shown, illustrating how both
the source data and the neutrino data lie inside, while the
instrumental backgrounds stay well outside.  We required all data
in the final signal sample to lie within the box shown in 
Figure~\ref{fig:cerbox}.

   The reconstruction quality cuts and the `Cerenkov box' cuts 
contribute to the overall acceptance loss, and we measured the scale of this 
loss along with the losses by the low level instrumental background cuts
by using Cerenkov and laser calibration sources.  The systematic 
uncertainties on these losses are associated with the calibration
sources themselves (source reflectivity and shadowing), the low level
electronic calibrations (for example, ADC pedestals), and changes in the
detector over time.  Using the calibration source data, we find that the total 
loss for all cuts is $1.4^{+0.7}_{-0.6}$\%.

	For the final signal sample, we further restricted events to be
within a fiducial volume of 550~cm and have a kinetic energy $T>6.75$~MeV.
The fiducial volume restriction minimizes backgrounds associated with
the acrylic vessel, light water, and PMTs, while the energy threshold
reduces radioactive backgrounds and neutron events in the final signal
sample.
\begin{table}
\caption{\label{tbl:events}  Data processing steps.}
\begin{tabular}{|lr|}
\hline
Analysis step & No. of events \\
\hline
Total event triggers   &    355~320~964  \\
Neutrino data triggers &    143~756~178     \\
$N_{\rm {hit}}\geq$30  &    6~372~899 \\
Inst. bkgrnd cuts  &  1~842~491     \\
Muon followers  &  1~809~979 \\
Cerenkov box cuts   &  923~717 \\
Fiducial volume cut  & 17~884 \\
Threshold cut        & 1 169  \\
\hline
Total events        &  1 169 \\
\hline
\end{tabular}
\end{table}    
	Table~\ref{tbl:events} summarizes the data processing, from the
total number of events in the raw data set to the final sample of 1169
events.  The table also includes the effects of cuts which remove the
spallation products of cosmic ray muons.  These cuts remove all events
within 20 s of a parent muon.  At this stage of the analysis, we have a data 
set which has been reconstructed and calibrated, and has had the majority of 
backgrounds removed.  However, before fitting the resulting distributions, we 
still need to build a model of the detector's response, and demonstrate that 
the background removal has been successful enough that we can perform the fits 
using a signal-only model.

\subsection{Model Building}

	The model of detector response we have used in this analysis 
takes as its inputs the physics of electron and $\gamma$ interactions in
matter, the geometry of the detector, the behavior of the front-end
data acquisition electronics and trigger, and---most importantly---the
same measured optical parameters used in the energy calibration described
in the previous section.  The model is a Monte Carlo simulation, which
combines these inputs as well as the state of the detector as a function
of time (the number of channels online, the overall energy scale determined
by the \NS~source, etc.) to produce a predicted response function for
all event positions, directions, and energies.  This response function
is what we use to create the pdfs which are ultimately used to fit the
data.

	To ensure that the model is correct, we tested it against 
Cerenkov data representative of the neutrinos we are trying to detect, for as 
many positions, directions, and energies as possible.  The degree to which the 
model does not correctly reproduce the various measurements sets the scale for 
the systematic uncertainties on the predicted response function.

\begin{figure}
\epsfxsize130pt
\figurebox{120pt}{160pt}{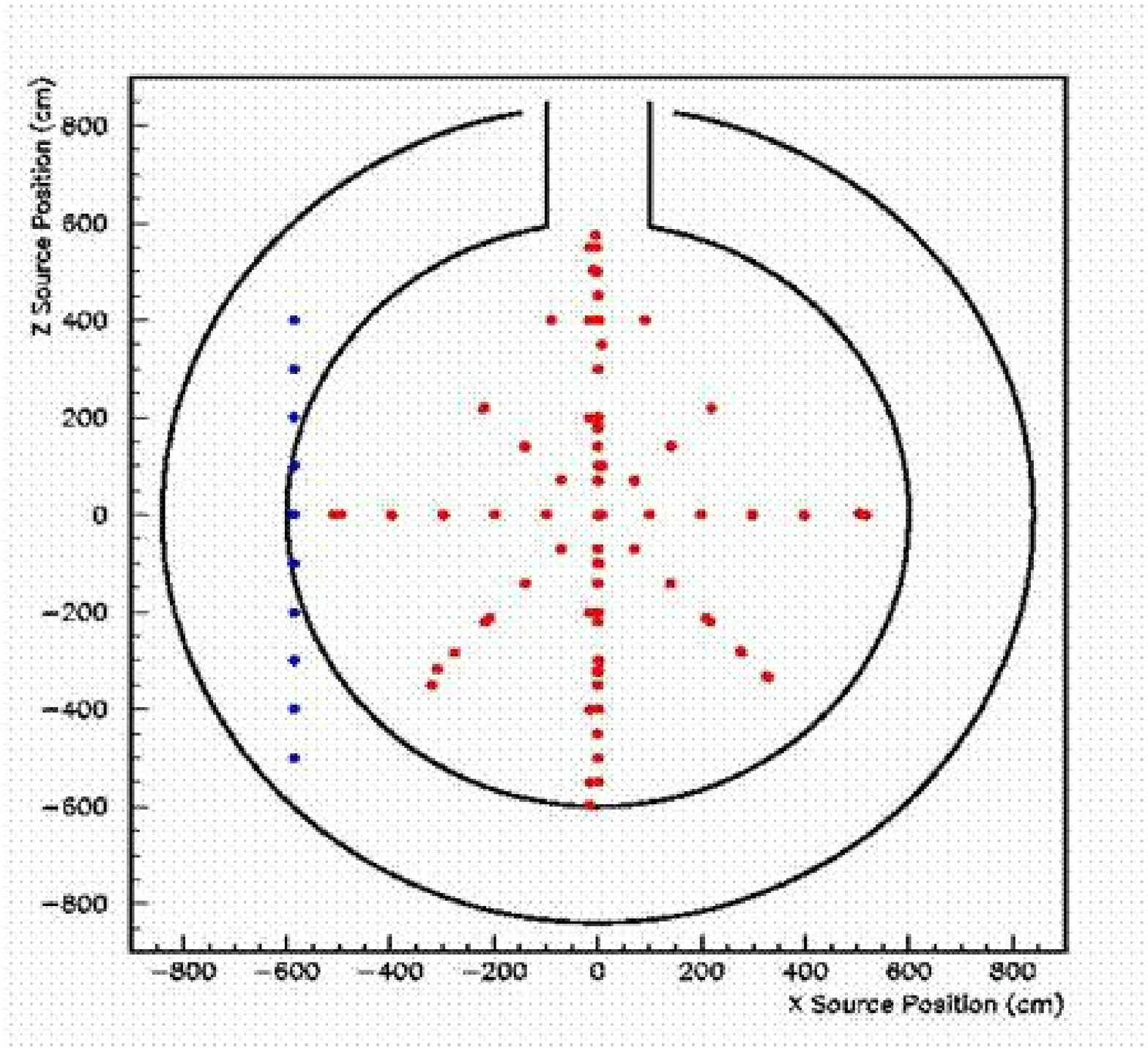}
\caption{Deployment positions for \NS~source for a \DO~scan and an \HO~scan.
\label{fig:n16pos}}
\end{figure}

\begin{figure*}
\epsfxsize30pc
\figurebox{16pc}{32pc}{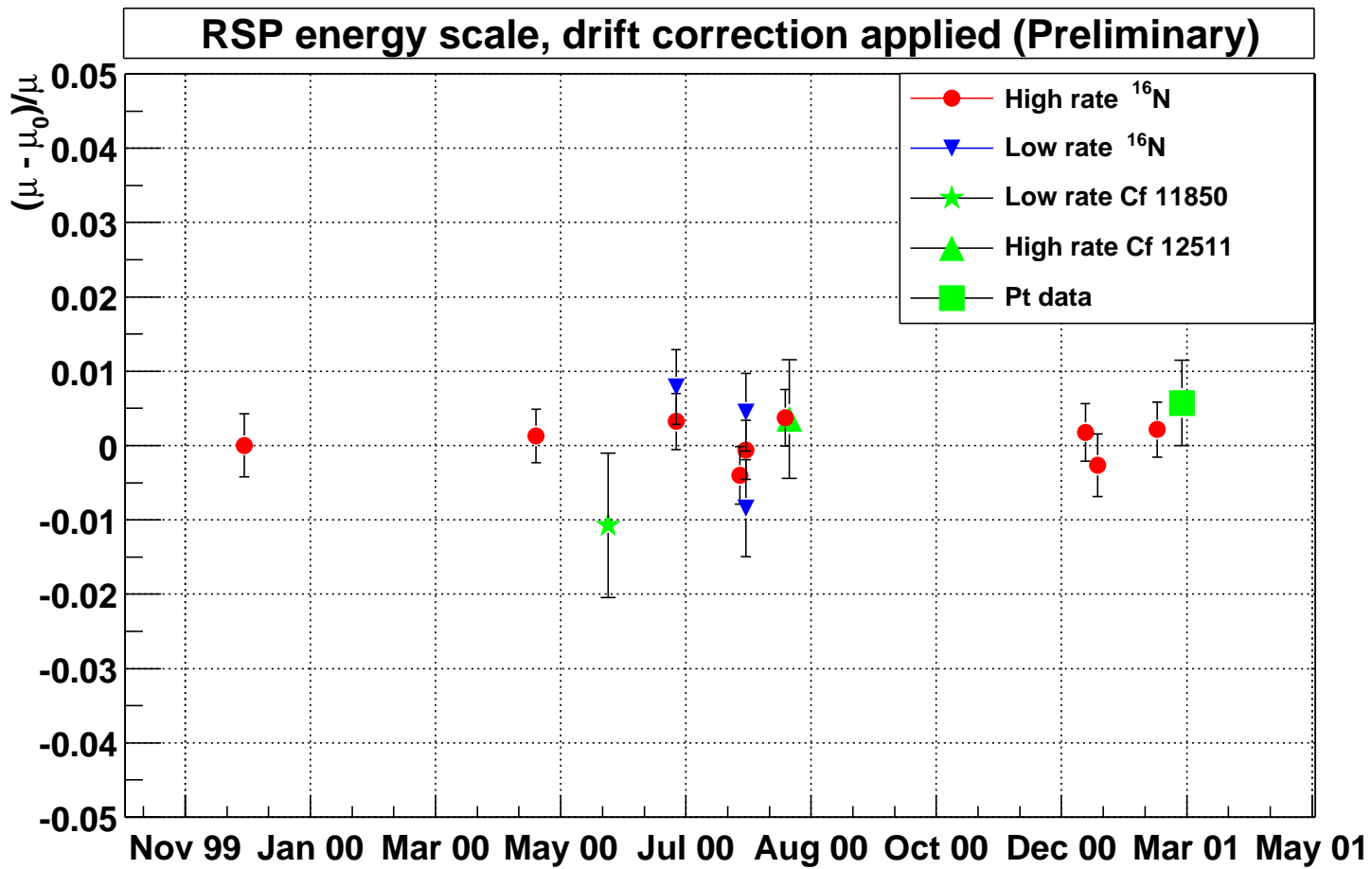}
\caption{Differences between predicted energy response and measured response
for different sources as a function of time.  In addition to the 
\NS~source, are a 19.8~MeV $\gamma$ from the pT source and a 
6.25~MeV $\gamma$ from the neutrons produced by the Cf source.\label {fig:ensump}}
\end{figure*}

	Figure~\ref{fig:n16pos} depicts the positions inside the
\DO~and \HO~for some of the \NS~scans.  For the dependence on energy of
the energy response, we compared \NS~data to pT data (6.13~MeV $\gamma$'s to 
19.8~MeV $\gamma$'s).  For the dependence on position and direction we
compared different source
positions and different sources---the Cf neutron source, for example, provides
a very different event position distribution than the \NS~source does, and
samples many more positions within the volume.  We also tested the
dependence on data rate by varying the rates for some of the 
calibration sources.  Figure~\ref{fig:ensump}
summarizes the differences between the predicted energy response and the
measured response for various sources as a function of time.  The overall
systematic uncertainty on the energy scale determined through these 
measurements is $1.4$\%.

\begin{figure}
\epsfxsize130pt
\figurebox{130pt}{160pt}{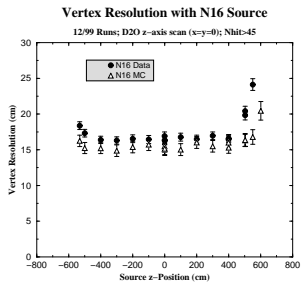}
\caption{Differences between predicted and measured vertex resolution
for the \NS~source as a function of position in the detector.
\label {fig:vres}}
\end{figure}

	We performed the same kinds of tests for the prediction of the 
reconstruction accuracy, and Figure~\ref{fig:vres} compares the vertex 
resolution measured with the \NS~source at various positions to the model 
prediction.  There is a small systematic shift ($\sim$ 1~cm) between the two, 
but otherwise the model tracks the data well.  The model prediction of the 
angular resolution agrees very well with the measurements made with the
\NS~source, and has a negligible contribution to the overall systematic 
uncertainty on the measured fluxes.

\subsection{Backgrounds}

	We are not quite ready yet to fit for the signal amplitudes, because
we must still demonstrate that the data is free enough from backgrounds to
justify the use of a model which contains only signal distributions. There
are three classes of background: the instrumental backgrounds discussed in 
Section~\ref{sec:dproc}, high energy $\gamma$ rays from the phototube support 
sphere and cavity walls, and low energy backgrounds from radioactivity both 
within and without the \DO~volume.  

	To measure the residual instrumental backgrounds, we used the
`Cerenkov box' cuts described earlier.  The low level cuts
aimed at reducing the instrumental backgrounds and the higher level
Cerenkov box cuts are independent and orthogonal---and so the fraction of the 
instrumental backgrounds which lie inside the Cerenkov box is the same whether 
the instrumental background cuts identify the events or not.  We therefore
used the fraction of identified instrumental backgrounds which lie inside
the box and multiplied it by the number of events in the `clean' data
sample which lie outside.  From this, we found the fraction of the clean 
data sample inside the Cerenkov box which may be due to instrumental 
backgrounds missed by the low level cuts.  As a fraction of the final CC data 
sample, this is $< 0.2$\%, small enough to ignore in the fit for signals.

	The determination of the high energy backgrounds was similar
in principle, but here we had at our disposal calibration sources which
provide triggered samples of high energy $\gamma$ rays.  The \NS~source
is nearly ideal for this measurement, as it acts as a triggered
`point source' of events which---with the exception of energy spectrum---look 
exactly like the background we are trying to measure.  To use this
source to measure the backgrounds, we deployed it near where the backgrounds
originate---out (and beyond) the detector's active volume.   We then
measured the ratio of the number of inward-going $\gamma$ events reconstructing 
just inside the source position (the `monitoring' box) to the number of events 
reconstructing inside the 550~cm fiducial volume.  With the number of events 
in the final data sample which reconstruct inside the same (but now 
spherically symmetric) monitoring box, we determined the number of background
events which lie insde the 550~cm volume by multiplying 
by the source-measured ratio.  We explored systematic uncertainties by varying 
the monitoring box size, the deployment position, and by using Monte Carlo 
simulation to explore the variation of the leakage with energy.  The final 
limit on this source of background measured in this way is $<$ 0.8\%.

\begin{figure}
\epsfxsize130pt
\figurebox{120pt}{160pt}{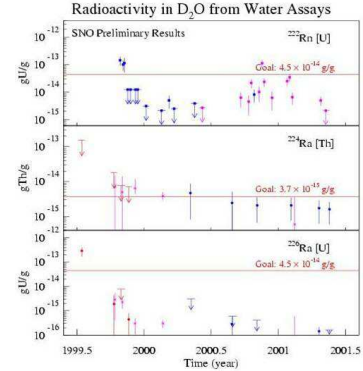}
\caption{Water assay measurements of radioactivity in \DO. \label {fig:assays}}
\end{figure}

Low energy backgrounds originate from several sources: radioactivity in the 
heavy water, the acrylic, the light water, and the PMTs.  Their typical energy 
is $\sim$ 2~MeV, and our energy threshold of 6.75~MeV is high enough that the 
leakage can only come from the tail of the background energy spectrum.  The
small fiducial volume of 550~cm also greatly restricts the number of
events from the PMTs, light water, and acrylic.  To estimate the number of 
events from low energy backgrounds which
leak above the signal energy threshold or inside the fiducial volume, 
we used a combination of radioassays, encapsulated U and Th 
calibration sources, and Monte Carlo simulation.  Figure~\ref{fig:assays} shows
that the radioactivity  in the heavy water---as determined by radioassays---is 
well below the original target values.  At these levels, simulation shows that 
the tail of the backgrounds above our energy threshold is negligible.  In the 
light water, assays also 
show that the the backgrounds are near or below target levels, but because 
these levels are still relatively high, we deployed calibration sources to 
measure the fraction that reconstruct within the 550~cm fiducial volume.
Of the external sources of background, by far the largest is the radioactivity
in the PMTs themselves.  With calibration sources placed near the PMT sphere,
we measured an upper limit on the leakage of the PMT radioactivity of 
$< 0.2$\% of the final CC rate.

\section{Results and Implications}

	We now have satisfied all the pre-requisites for doing a signal
extraction: we have a clean data set in which the backgrounds are low enough 
to justify a signal-only fit, and a model which correctly predicts the 
response function of the detector as measured by calibration sources.
For the neutrino spectrum input to the model, we use an undistorted 
$^8$B shape~\cite{ortiz,hep}.

\begin{figure*}
\epsfxsize20pc
\figurebox{16pt}{32pt}{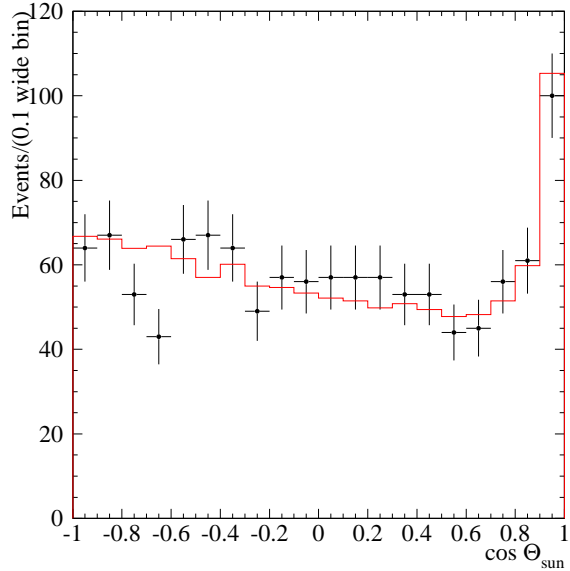}
\caption{Reconstructed direction distribution of electrons
with respect to the Sun. \label {fig:cstsun}}
\end{figure*}

	The maximum likelihood fit to the 1169 events in our sample
gives us 975.4$\pm$39.7 CC events, 106.1$\pm$15.2 ES events, and 
87.5$\pm$24.7  neutron events, where the uncertainties given are statistical 
only.  Figure~\ref{fig:cstsun} shows the best fit to the distribution
of event directions with respect to the Sun. The elastic scattering
peak can clearly be seen, but with the available statistics, only a hint of 
the slope of the CC electrons.

	To convert the CC and ES event numbers into fluxes, we need to correct
for the acceptance of the cuts, the energy threshold, and the fiducial
volume restriction.  We then need to normalize by the interaction
cross sections and the number of deuterons and electrons inside the fiducial
volume.  For the CC cross section, we use the calculation of 
Butler~{\it et al}~\cite{xsect}, and do not include any radiative 
corrections.  The radiative corrections may serve to increase the cross 
section by up to a few percent~\cite{rad_corr}, and therefore {\it decrease} 
the measured value of the flux (and ultimately increase the significance of any 
difference between the CC and ES fluxes).  We also include small corrections 
due to the isotopic abundances of $^{17}$O and $^{18}$O, upon which CC 
reactions can also occur.  Finally, we normalize by the overall livetime.

	Table~\ref{tbl:sysuncs} lists the systematic uncertainties on the
flux measurements.  The dominant uncertainties arise from our lack of
knowledge of the true response function of the detector.  As described
above, we characterize the scale of the uncertainties on the model by
comparing the model predictions to measurements made with calibration sources,
for example the 1.4\% on the energy scale.  To derive the uncertainties on the 
fluxes shown in Table~\ref{tbl:sysuncs}, we varied the model predictions
over the range of the uncertainties and repeated the analysis.  In
some cases this resulted in a larger uncertainty on the flux measurement---the
1.4\% uncertainty on the energy scale becomes a $\sim$ 6\% uncertainty
on the flux derived from the CC rate, for example.
\begin{table}
\caption{\label{tbl:sysuncs} Systematic uncertainties on fluxes.}
\begin{center}
\begin{tabular}{|lrr| }
\hline
Error source  & CC  error  & ES  error\\
                & (percent) & (per cent)\\
\hline
Energy scale            & -5.2, +6.1  & -3.5 ,+5.4 \\
Energy resolution       & $\pm$0.5       & $\pm$0.3       \\
En. non-linearity           & $\pm$0.5       & $\pm$0.4 \\
Vertex accuracy         & $\pm$3.1       & $\pm$3.3       \\
Vertex resolution       & $\pm$0.7       & $\pm$0.4       \\
Angular resolution      & $\pm$0.5       & $\pm$2.2       \\
High energy $\gamma$'s  & -0.8, +0.0     & -1.9, +0.0  \\
Low energy bkgrnd   &  -0.2, +0.0     & -0.2, +0.0 \\
Inst. bkgrnd 		&  -0.2, +0.0    & -0.6, +0.0 \\
Trigger efficiency      &  0.0           & 0.0 \\
Live time               & $\pm$0.1       & $\pm$0.1       \\
Cut acceptance          & -0.6, +0.7     & -0.6, +0.7  \\
Earth orbit ecc. & $\pm$0.1      & $\pm$0.1 \\
$^{17}$O, $^{18}$O      &  0.0           &  0.0 \\
\hline
Exp. uncertainty  & -6.2, +7.0     & -5.7, +6.8  \\
\hline
Cross section           & 3.0   & 0.5     \\
\hline
Solar Model             & -16, +20  & -16, +20  \\
\hline
\end{tabular}
\end{center}
\end{table} 

	In addition to the measurement of the systematic uncertainties,
we have explored the systematic behavior of our results under many different
analysis approaches: for example, comparing different suites of low level cuts, 
reconstruction algorithms, Cerenkov box cuts, and choices of fiducial volume.  
We have also compared the results we get using the total number of hit
PMTs as the measure of energy scale (thus changing the sensitivity to the
knowledge of the late light distribution) to the results from the
energy calibration described above. Lastly, we have performed fits using
an analytical (as opposed to Monte Carlo) model of the detector
response.  In all cases, the results from these alternative approaches
agree with the fluxes presented here to well within the systematic uncertainties
quantified in Table~\ref{tbl:sysuncs}.

	Converting the fit numbers to fluxes and including the systematic
uncertainties listed in Table~\ref{tbl:sysuncs}, we find that the flux of 
neutrinos inferred from the ES reaction (assuming no flavor transformation) is 
\begin{eqnarray*}
\phi_{\rm{SNO}}^{\rm{ES}}(\nu_x) & = & 2.39 \pm 0.34 ({\rm stat.}) \\
  &&    ^{+0.16}_{-0.14}~({\rm sys.}) \times 10^6~{\rm cm}^{-2} {\rm s}^{-1}
\end{eqnarray*}
and the flux of $^8$B $\nu_e$'s measured by the CC reaction is  
\begin{eqnarray*}
\phi_{\rm{SNO}}^{\rm{CC}}(\nu_e) & = & 1.75 \pm 0.07~({\rm stat.}) ^{+0.12}_{-0.11}~({\rm sys.}) \\ 
			     &&    \pm 0.05~({\rm theor.}) 
                               \times 10^6~{\rm cm}^{-2} {\rm s}^{-1} \\
\end{eqnarray*}
where the theoretical uncertainty comes from the uncertainty in the
CC cross section~\cite{xsect}.
The difference between these two numbers is $1.6 \sigma$, assuming that
the systematic errors are distributed normally.  The low significance of
this result is driven mainly by the large statistical errors on the
ES measurement.  However, the Super-Kamiokande collaboration has measured
the flux with the ES reaction to high precision~\cite{SK}, and finds 
\begin{eqnarray}
\phi_{\rm{SK}}^{\rm{ES}}(\nu_x) & = & 2.32 \pm 0.03~({\rm stat.})
^{+0.08}_{-0.07}~({\rm sys.}) \\
&& \times 10^6~{\rm cm}^{-2} {\rm s}^{-1}.
\end{eqnarray} 
The difference between SNO's measurement
using the CC reaction (sensitive only to $\nu_e$'s) and 
Super-Kamiokande's measurement using the ES reaction (sensitive to $\nu_{\mu}$'s
and $\nu_{\tau}$'s as well as $\nu_e$'s in the ratio of 1./6.5) is
$3.3 \sigma$.  This difference is therefore evidence of an active, 
non-$\nu_e$ component to the solar $^8$B neutrino flux.

\begin{figure}
\epsfxsize130pt
\figurebox{120pt}{160pt}{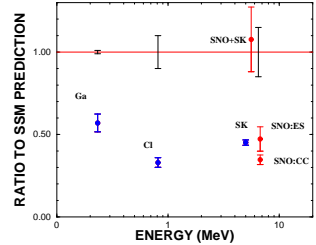}
\caption{Summary of solar neutrino rate measurements from various experiments
relative to BPB01 standard solar model, including SNO and the derived 
$^8$B flux from the SNO and Super-Kamiokande rates. \label {fig:ssm_sno}}
\end{figure}
	Figure~\ref{fig:ssm_sno} summarizes the situation for all published
solar neutrino experiments, including SNO.  The points are plotted as
ratios of the measured fluxes to the Standard Solar Model predictions
of Bahcall, Pinnsoneault and Basu (BPB01~\cite{BPB}), for the energy threshold 
used in each experiment.  Here we can see the $3.3\sigma$ difference between 
the SNO and Super-Kamiokande
measurements as well as the poor statistical accuracy of the SNO ES.  
Also plotted in Figure~\ref{fig:ssm_sno} is the total flux of all 
$^8$B neutrinos using the SNO CC measurement and the Super-Kamiokande ES 
measurement:  
\begin{displaymath}
\phi(\nu_x) =  5.44\pm 0.99 \times 10^6~{\rm cm}^{-2}
{\rm s}^{-1}.
\end{displaymath}
We see in the figure that the agreement between the 
measurement and the model prediction is very good.

\begin{figure*}
\epsfxsize20pc
\figurebox{16pc}{32pc}{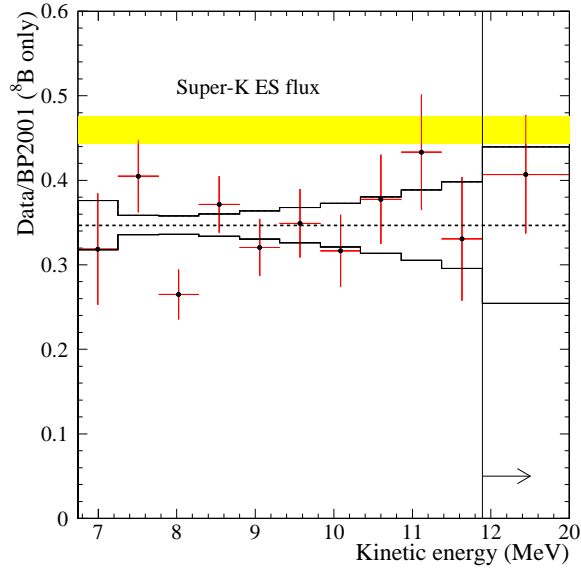}
\caption{Spectrum of electrons from $\nu_e$ interactions compared to
prediction of the BPB01 standard solar model. \label{fig:espec}}
\end{figure*}

	Figure~\ref{fig:ssm_sno} also shows that the differences
in the thresholds for the SNO and Super-Kamiokande measurements allows
for the possibility that there is some spectral distortion which could
be causing the difference.  Such a spectral distortion could occur if,
for example, the oscillation were into a sterile neutrino.  To look for
such an effect, we can first examine the spectrum of recoil electrons
created by the CC interactions relative to the prediction for an
undistorted $^8$B spectrum.  We derive such a spectrum by re-fitting
the data energy bin-by-energy bin, without using the pdf for the CC energy 
spectrum.  Figure~\ref{fig:espec} shows the ratio of the spectrum derived
this way to the standard solar model prediction.  The dominant systematic  
uncertainties are indicated by the horizontal lines on the plot.  
Figure~\ref{fig:espec} shows that there is no large distortion in the
expected spectrum.

We can also eliminate the possibility of a spectral distortion leading to
the difference in the SNO CC and Super-Kamiokande ES measurements by comparing
the two measurements for the same neutrino energy.  As described by
Fogli~{\it et al}~\cite{fogli}, this can be done by using different recoil
energy thresholds for the SNO and Super-Kamiokande measurement.  For these
`matched' thresholds ($\sim 8.5$~MeV for the Super-Kamiokande measurement
compared to SNO's 6.75~MeV)  we still get a difference of $3.1\sigma$.  This 
measurement is independent of both the standard solar model flux prediction 
and the predicted shape of the $^8$B spectrum---we need only know that the 
Sun produces $\nu_e$'s to see that there is a change to other active flavors.

\section{Future and Conclusions}

SNO's current and future data sets will provide many more
interesting measurements.  We are now analyzing the pure
D$_2$O data in order to make a the measurement of the NC rate.  The NC
measurement should give us a confirmation of the CC-ES result, a higher 
precision measurement of the total $^8$B flux, and a higher significance for 
the excess of non-$\nu_e$ flavors (because
we will be comparing a $\nu_e$ flux of 
$1.75 \times 10^6 {\rm cm}^{-2}{\rm s}^{-1}$ to 
$\sim 5 \times 10^6 {\rm cm}^{-2}{\rm s}^{-1}$ rather than to
$\sim 2.3 \times 10^6 {\rm cm}^{-2}{\rm s}^{-1}$).  We are also analyzing
the data in day and night bins, to determine whether any asymmetry
is present (which would indicate that matter oscillations are the
cause of the flavor change as well as better
restrict the allowed regions in the ($\tan^2\theta$,$\Delta m^2$) plane.
In addition, we are working on an analysis which includes the hep neutrinos.

	Beyond the pure D$_2$O data, we will also have the salt data
set, which should provide us with an even better NC measurement, as well
as new measurements of the other fluxes as well.  Non-solar
neutrino physics analyses are underway as well---looking at atmospheric 
neutrinos, anti-neutrinos (for which SNO has an exclusive coincidence tag),
and supernova searches. 

	SNO's first results, in combination with the Super-Kamiokande 
collaboration's measurements, provide direct evidence that solar neutrinos
undergo flavor change on their way from the Sun to the Earth.  They also
show that the Standard Solar Model prediction of the $^8$B flux is correct
within the uncertainties of both the prediction and the measurement.  We
expect many more interesting measurements to come out of SNO for a long
time to come.  

\section*{Acknowledgments}
This research was supported by the Natural
Sciences and Engineering Research Council of Canada, Industry Canada, National
Research Council of Canada, Northern Ontario Heritage Fund Corporation
and the Province of Ontario,  the United States Department of
Energy, and in the United Kingdom by the Science and Engineering
Research Council and the Particle Physics and Astronomy Research
Council.  Further support was provided by INCO, Ltd., Atomic Energy of
Canada Limited (AECL), Agra-Monenco, Canatom, Canadian
Microelectronics Corporation, AT\&T Microelectronics, Northern Telecom and 
British Nuclear Fuels, Ltd.   The heavy water
was loaned by AECL with the cooperation of Ontario Power Generation.


\begin{thebibliography}{99}
\bibitem{hep} Given the limit set for the {\it hep} flux by
Ref.~\cite{SK}, the effects of the {\it hep} contribution may
increase this difference by a few percent.
\end{thebibliography}
\end{document}